\documentclass{aastex61}

\newcommand\aastex{AAS\TeX}

\usepackage{bm}
\usepackage{color}

\received{}
\revised{}
\accepted{}
\submitjournal{ApJ}

\shorttitle{\aastex\ On the Impact Origin of Phobos and Deimos I\hspace{-.1em}I}
\shortauthors{Hyodo et al.}

\begin{document}
\title{On the Impact Origin of Phobos and Deimos I\hspace{-.1em}I:\\True Polar Wander and Disk Evolution}

\correspondingauthor{Ryuki Hyodo}
\email{hyodo@elsi.jp}

\author{Ryuki Hyodo}
\affil{Earth-Life Science Institute/Tokyo Institute of Technology, 2-12-1 Tokyo, Japan}
\affiliation{Institut de Physique du Globe/Universit{\'e} Paris Diderot/CEA/CNRS, 75005 Paris, France}

\author{Pascal Rosenblatt}
\affiliation{ACRI-ST, Nice, France (current)}
\affiliation{Royal Observatory of Belgium, 1180 Brussels, Belgium}

\author{Hidenori Genda}
\affil{Earth-Life Science Institute/Tokyo Institute of Technology, 2-12-1 Tokyo, Japan}

\author{S{\'e}bastien Charnoz}
\affiliation{Institut de Physique du Globe/Universit{\'e} Paris Diderot/CEA/CNRS, 75005 Paris, France}



\begin{abstract}
Phobos and Deimos are the two small Martian moons, orbiting almost on the equatorial plane of Mars. Recent works have shown that they can accrete within an impact-generated inner dense and outer light disk, and that the same impact potentially forms the Borealis basin, a large northern hemisphere basin on the current Mars. However, there is no a priori reason for the impact to take place close to the north pole (Borealis present location) nor to generate a debris disk in the equatorial plane of Mars (in which Phobos and Deimos orbit). In this paper, we investigate these remaining issues on the giant impact origin of the Martian moons. First, we show that the mass deficit created by the Borealis impact basin induces a global reorientation of the planet to realign its main moment of inertia with the rotation pole (True Polar Wander). This moves the location of the Borealis basin toward its current location. Next, using analytical arguments, we investigate the detailed dynamical evolution of the eccentric inclined disk from the equatorial plane of Mars that is formed by the Martian-moon-forming impact. We find that, as a result of precession of disk particles due to the Martian dynamical flattening $J_{2}$ term of its gravity field and particle-particle inelastic collisions, eccentricity and inclination are damped and an inner dense and outer light equatorial circular disk is eventually formed. Our results strengthen the giant impact origin of Phobos and Deimos that can finally be tested by a future sample return mission such as JAXA's Martian Moons eXploration (MMX) mission.
\end{abstract}

\keywords{planets and satellites: composition, planets and satellites: formation, planets and satellites: individual (Phobos, Deimos)}



\section{Introduction} \label{sec:intro}
The origin of the two small Martian moons, Phobos and Deimos, is still debated. It has been believed that they were captured asteroids due to their spectral properties resembling those of D-type asteroids \citep[e.g.][]{Mur91,Bur78}. However, the captured scenario has been confronted with the challenge of explaining their almost circular equatorial orbits around Mars (inclinations of 1.08$^\circ$ and 1.79$^\circ$ from Mars' equator for Phobos and Deimos, respectively). In contrast, accretion within debris disk formed by a giant impact may naturally explain their orbital configurations \citep{Cra94,Cra11,Ida97,Hyo15a,Hyo15b} but \cite{CC12} showed that Phobos and Deimos can not be formed directly from the spreading of a ring interior to the Roche radius. Then, \cite{Ros16} have shown that these small moons can be formed by accretion within a thin debris disk that extends outside the Roche limit ($\sim 3 R_{\rm Mars}$ where $R_{\rm Mars}$ is the Mars' radius) and is sculpted by an outward migration of a large inner moon that is formed by the spreading of a thick disk lying inside the Roche limit. Furthermore, \cite{Hes17} have shown that the tidal disruption of such a large inner moon during the tidal decay creates a new generation of rings/disks around Mars followed by the spreading and accretion of smaller moons. They showed that such a process could have occurred repeatedly over the past 4.3 billion years, suggesting that Phobos is the last generation of moon we observe today. Note that, currently the only successful scenario to form Deimos is the accretion within an extended outer disk proposed by \cite{Ros16}.\\
 
Such a Martian-moon-forming disk can be created by a giant impact that can also form the asymmetric northern lowland that is the Borealis basin: impactor mass of $\sim 0.03 M_{\rm Mars}$ (about $1/3$ Martian radius) and an impact velocity of $\sim$ 6 km s$^{-1}$ \citep{Mar08,Cit15,Ros16,Hyo17c}. However, two natural questions arise. First, without or a slow pre-impact spin of Mars (compared to the impact spin angular momentum), a giant impact spins up Mars and thus a debris disk is generated symmetrically around the equatorial plane. In this case, the impact point (that is the Borealis basin) is expected to be located near the equator. So, why is the Borealis basin currently located on the northern hemisphere and not on the equatorial plane? Second, during protoplanet formation through successive accretion of planetesimals, a protoplanet may naturally have a rotation \citep{Oht98}. Thus, if Martian pre-impact spin is comparable and not aligned to the angular momentum delivered by the giant impact, the resultant debris disk is expected to be inclined with respect to the equatorial plane of Mars. Also, the location of the Borealis basin may not be in the polar region. So, why do Phobos and Deimos orbit almost on the Martian equatorial plane? And, why is the Borealis basin currently located on the northern hemisphere?  In addition, just after the giant impact, the orbits of the debris are eccentric (Section \ref{subsec:a_e}). Thus, we need a dynamical path to form a circular equatorial disk from which Phobos and Deimos can accrete \citep{Ros16}.\\

In section \ref{sec:tpw}, using analytical arguments, we show that the Borealis-basin-induced reorientation of the planet (True Polar Wander (TPW)) can explain the dynamical path of the Borealis basin settling in its current location in the Northern hemisphere. In section \ref{sec:disk_evo}, we detail the dynamical evolution of the initial disk that is formed by a Borealis basin forming impact and show that an initial ''inclined'' (with respect to Mars' equatorial plane) and eccentric disk will settle into a thin equatorial circular disk, forming equatorial orbiting Phobos and Deimos. In section \ref{sec:summary}, we summarize our results. \\

 \section{True Polar Wander}\label{sec:tpw}
The Borealis impact basin is expected to be located near the equator of the planet, if one assumes that the giant collision that caused it gave Mars most of its present spin. Even if Mars had a significant pre-impact spin, it is unlikely that the Borealis basin could form directly beneath the polar region. Under this assumption, one needs to explain the current polar position of the Borealis basin. Here, we check for True Polar Wander (TPW) as a mechanism for providing the required motion from the equator to polar areas.\\

\begin{figure}[ht!]
\epsscale{1.0}
\plotone{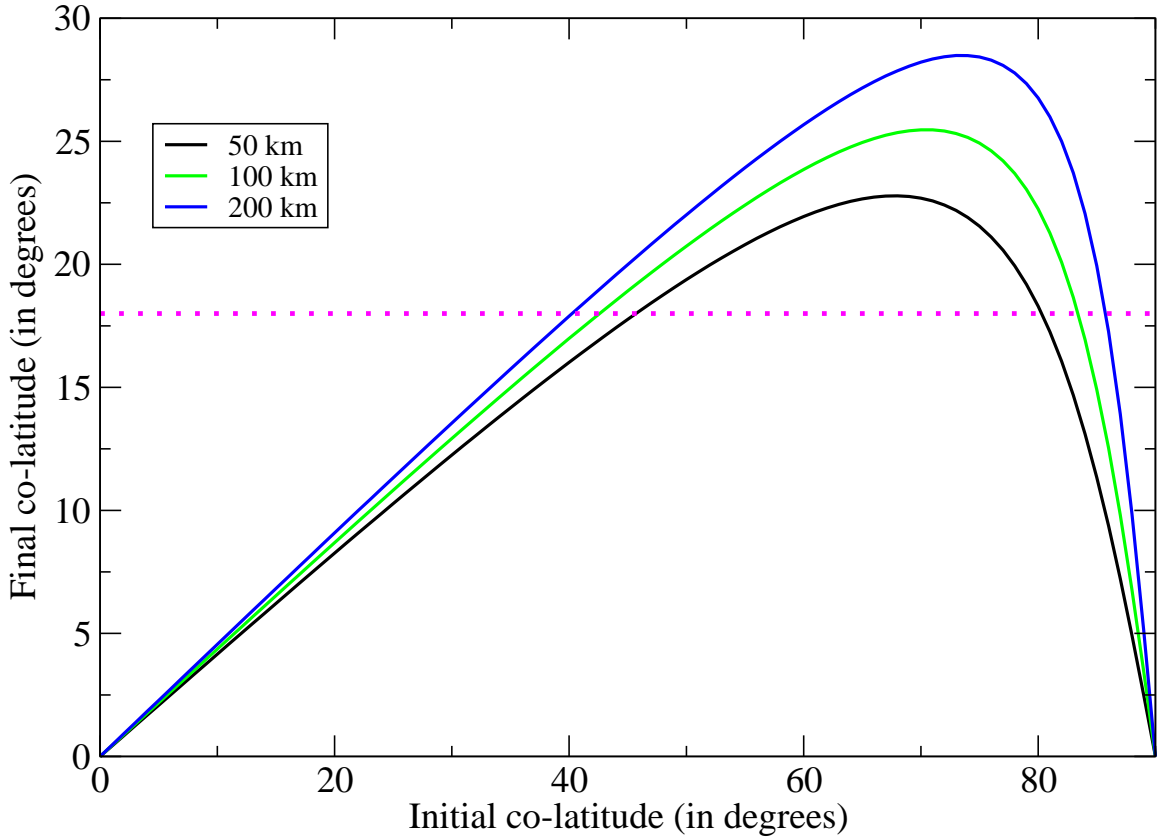}
\caption{Predictions of True Polar Wander (TPW) due to Borealis impact basin as a function of lithospheric thickness as labeled on the plot. The horizontal dotted line represents the pre-Tharsis TPW (or post-Borealis TPW) co-latitude of Borealis basin (see text).}
\label{tpw_res}
\end{figure}

We use the equilibrium theory proposed by \cite{Gol55} and adapted by \cite{Wil84}  and \cite{Mat06} for assessing the Mars True Polar Wander induced by the Tharsis bulge. This theory consists of computing the effect of surface mass excess load and the effect of the rotational bulge excess mass on the inertia tensor of the planet. The basic calculations of this theory are the diagonalization of the perturbed inertia tensor to provide the position of the new pole at the surface of the planet after TPW motion \citep[see also][]{Mat06}. In this theory, the latitudinal shift $\delta$ of the pole at the surface of the planet is given by the following relationship:
\begin{equation}
	\delta = \frac{1}{2}\tan^{-1} \left[ \frac{Q'\alpha \sin(2\theta_{\rm L})}{1-Q'\alpha \cos(2\theta_{\rm L})} \right],
\label{tpw_eq}
\end{equation}
where $\theta_{\rm L}$ is the colatitude of the center of the axisymmetric excess mass before TPW, $Q'$ is the ratio of the mass excess over the rotational bulge load and $\alpha$ is a combination of load and tidal Love numbers of the planet (see Appendix). Although this theory has been applied for Mars with surface mass excess (Tharsis), it can also be applied to mass deficit such as that created by impact basins \citep{Wil84}. In that case, the center of the impact basin is expected to move from its initial location toward the pole of the hemisphere. Thus it is likely that the center of the Borealis was initially located in the northern hemisphere in order to move toward its current location. Then, $Q’$' has a negative value and is about $-1.102$ for the Borealis basin, depending on the tidal Love numbers of Mars, i.e. on the lithospheric thickness (see Appendix). Figure \ref{tpw_res} displays the expected position of the center of the Borealis basin after TPW (final position) versus its initial position.\\

However, we need to know the Borealis position after Borealis-TPW to find which pre-Borealis-TPW positions are suitable from Figure \ref{tpw_res}.  As Tharsis TPW is the most recent large TPW event in Mars history, we need to compute the position of Borealis before Tharsis TPW. The Tharsis-TPW displacement is toward the equator with amplitude of 18.9$^\circ$ \citep[see][for most recent estimations]{Mat10,Bou16}, meaning that Tharsis formed at higher latitudes by about 20$^\circ$ than it is today. This Tharsis-TPW has also moved the Borealis position. However, the present longitude is 208$^\circ$E \citep{And08}, which is off the Tharsis central meridian \citep[259.5$^\circ$E,][]{Mat10} by about 50$^\circ$. Simple spherical trigonometry considerations show that the TPW amplitude decreases when the distance to its central meridian increases. As the difference between Borealis and Tharsis central meridians is about 50$^\circ$, the Borealis center will not move by 18.9$^\circ$ in latitude but only by 5$^\circ$. It corresponds to a pre-Tharsis TPW co-latitude of 18$^\circ$ (latitude of 72$^\circ$) for Borealis center (see Appendix).\\

If one assumes no significant TPW events occur between Borealis and Tharsis-TPW ones, this 18$^\circ$ co-latitude of Borealis centre before the Tharsis TPW in turn suggests two possible initial (pre-borealis) latitude ranges between 45$^\circ$-50$^\circ$ and 5$^\circ$-10$^\circ$ (co-latitudes of 45$^\circ$-50$^\circ$ and of 85$^\circ$-80$^\circ$, respectively) for Mars' lithospheric thickness at time of TPW between 50 km and 200 km (see Figure \ref{tpw_res}). Intermediate latitudes, between 10$^\circ$N and 45$^\circ$N are possible for lithosphere thinner than 50 km. The 5$^\circ$-10$^\circ$ range of initial latitudes indicates large TPW displacement between -62$^\circ$ and -67$^\circ$ (3 times larger than Tharsis-TPW), so a Borealis initial position near the equator in agreement with a collision having given Mars most if its spin. The 45$^\circ$-50$^\circ$ range implies a more modest TPW between -22$^\circ$ and -27$^\circ$ (see Eq. 1) comparable to the absolute value of the Tharsis-TPW. The 45$^\circ$-50$^\circ$ range also implies that the giant collision would not have given to Mars most of its spin, meaning in turn that Mars would have significant pre-impact spin. This spin is however difficult to compute since it also depends on the angle and the kinetic energy of the collision.\\

The linear theory of \cite{Mat10} assumes elastic rheology for the lithosphere and the planet and so instantaneous TPW displacement.\cite{Cha14} have introduced a more realistic rheology allowing for assessing the timescale of the TPW displacement. They found a typical time scale of 20 Myr for the Tharsis TPW to reach the displacement predicted by the equilibrium theory (see Figure 4 in their paper), depending on the viscosity of the Martian mantle. At the time of Borealis formation the viscosity of the mantle is expected to be lower than at time of Tharsis formation, hence implying even shorter time scale for the Borealis-TPW displacement to take place. Thus, the Borealis center can easily reach its final position before the Tharsis-TPW takes place.\\

Although, the two solutions for Borealis-TPW are mathematically possible, the solution with lower TPW displacement seems more plausible than the larger one. Large TPW is indeed expected to produce planetary scale stresses in the lithosphere, which should be seen as a global tectonic pattern. But such global pattern is not observed at Mars surface today \citep[see][]{Bou16}. Although, these authors argue that the lithospheric stresses should be relaxed with time, the tectonic pattern should be still recorded in the lithosphere since the secular cooling of the planet stiffens the lithosphere with time. Actually, even modest TPW generates stresses in the lithosphere. But as these stresses are more modest, they may less likely produce tectonic pattern than in the case of huge TPW. Therefore, a more modest TPW is more in agreement with the absence of TPW-induced global tectonic pattern. Such issue deserves however more thoroughly investigations that will be performed in future works. In turn, the dichotomy might not be initially lying at the pre-Tharsis equator, in contrary to the hypothesis in \cite{Bou16}. Nevertheless, large impacts occurring between Borealis and Tharsis events might have produced TPW of a couple of degrees. Considering a few large impacts \citep{Bot17}, we indeed found that Utopia and Hellas could have produced such TPW, which in turn could have modified the position of Borealis center after its own TPW wrt the position derived from the Tharsis-TPW alone (so the estimation of the initial position of Borealis). Hence, thorough investigations are needed to fully draw Mars TPW history in order to check whether our Borealis TPW calculations are still in agreement with Tharsis TPW predictions within the error bar of the Tharsis center determination \citep{Mat10} and the error bar of the estimation of Tharsis TPW displacement from surface observations \citep[e.g.][]{Per07,Kit09}. This will be performed in future works.\\

The application of TPW linear theory suggests two possible solutions for Borealis-TPW: on one hand, large TPW and so near-equatorial position for Borealis impact centre and on the other hand, smaller TPW with $\sim 45^\circ$ latitude impact center. The latter solution is however favoured for physical considerations, and in turn implies a non-equatorial debris disk. The evolution of this non-equatorial disk is studied in the next section.

\section{Dynamical evolution of the disk}
The results of TPW in Section \ref{sec:tpw} imply that the giant impact is more likely to occur not on the equatorial plane but at higher latitude. This indicates that the mean inclination of the debris just after the impact is expected to be inclined with respect to the equator, depending on the pre-impact spin state. In addition, the orbits of the initial debris are expected to be eccentric (see Section \ref{subsec:a_e}). However, in order to form Phobos and Deimos in the equatorial plane in a framework of the giant impact hypothesis, a dense inner equatorial and a light outer equatorial disk are required to form \citep{Ros16}. Inelastic collision can decrease eccentricity through energy damping but it conserves the angular momentum. The inner dense and outer light surface density profile of the Martian-moon-forming disk \citep[][and see their Figure 1]{Ros16} was generated as a consequence of the angular momentum conservation from the same $a-e$ distribution of the debris as that shown in Figure \ref{ae_dis} by calculating  an equivalent circular orbit of radius $a_{\rm eq}$ for every disk particles ($a_{\rm eq}=a(1-e^2)$) as done also in the case of the Moon-forming impact and its disk \citep[e.g.][]{Can04}. However, the detailed collisional evolution and its timescale are unclear. In this section, we address a dynamical path that can potentially bring the impact-generated inclined and eccentric debris to the circular equatorial disk that can form Phobos and Deimos.

\label{sec:disk_evo}
\subsection{Orbital elements of disk particles just after the impact}\label{subsec:a_e}
\begin{figure}[ht!]
\epsscale{0.6}
\plotone{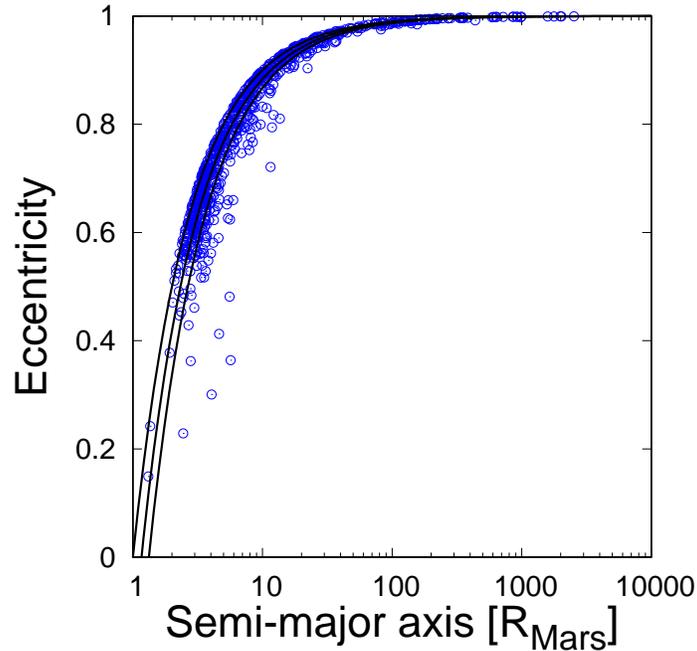}
\caption{Distribution of disk particles on semi-major axis to eccentricity plane in the case of the Borealis basin forming canonical impact (impact angle of 45 degrees with respect to the Martian surface and $N_{\rm SPH}=3 \times 10^6$ at $T=5$ h, where $N_{\rm SPH}$ is the number of SPH particles and $T$ is the time after the start of simulation). Data are taken from \cite{Hyo17c}. Here, we plot only particles whose pericenter distances are larger than the radius of Mars (disk particles). The solid black lines from top to bottom represent equation \ref{e_a_analytic} with $R_{\rm Mars}=3260$ km and $\Delta r = 0$, $0.5$ and $1.0R_{\rm imp}$, respectively taken from the simulation \citep{Hyo17c}, where $R_{\rm imp} \sim 1000$ km is the radius of the impactor.}
\label{ae_dis}
\end{figure}

Figure \ref{ae_dis} shows the distribution of eccentricity of disk particles just after the impact, which is taken from the Martian moon-forming impact simulation in \cite{Hyo17c}. We find that the disk particles initially have a wide distribution of large eccentricity (up to almost 1) around Mars. The distribution of semi-major axes and eccentricities of disk particles just after the impact can be analytically derived as follows. For particles that end up orbiting around Mars (called “disk particles”), the near impact point is expected to be their pericenters (because Keplerian orbits are closed trajectories, and if their pericenter is inside Mars, they would be accreted very rapidly). Therefore, pericenter distances of the disk particles can be written as $r_{\rm peri} \sim R_{\rm Mars}+\Delta r = a(1-e)$ where $\Delta r$ is the small distance from the surface of Mars, $a$ is the semi-major axis and $e$ is the eccentricity. Thus, we can derive the relationship between $a$ and $e$ as
\begin{equation}
	e = 1 - \frac{R_{\rm Mars} + \Delta r}{a}.
\label{e_a_analytic}
\end{equation}
Thus, the distribution of $a$ and $e$ of disk particles just after the impact is expected to be along the line obtained by equation \ref{e_a_analytic} (Figure \ref{ae_dis}), and this means that all particles share almost the same pericenter distances.\\

\subsection{Orbital evolution of disk particles}\label{sec:dynamics}
\begin{figure}[ht]
\epsscale{0.6}
\plotone{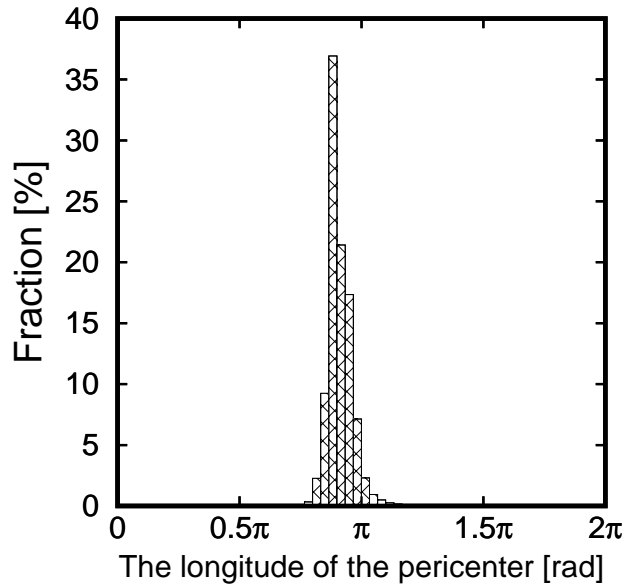}
\caption{Distribution of the longitude of the pericenter in the case of the Borealis basin forming canonical impact ($N_{\rm SPH}=3 \times 10^6$ at $T=5$ h obtained from \cite{Hyo17c}, see also their Figure 1).}
\label{phases}
\end{figure}

Just after the impact, the disk particles are almost aligned (phase alignment) at their longitude of the pericenters (Figure \ref{phases}) with wide distributions of eccentricity and semi-major axis (Figure \ref{ae_dis}). In addition, the disk may be initially inclined from the Martian equatorial plane depending on the giant impact condition. Thus, under these circumstances, collision velocities between nearby particles are significantly small about their shear motion and thus collisional damping is not effective, unless collision happens at pericenter between particles whose orbital elements are significantly different. In this subsection, we will discuss two extreme cases of the expected dynamical evolutions of the system where the system forms a torus-like structure and where the system damps very quickly to form a thin inclined nearly circular disk. Note that, direct $N$-body simulations including fragmentation is necessary to understand more details about the disk evolution. However, this is beyond the scope of this work and we will leave this matter in the future work.

\subsubsection{The case for forming a torus-like structure}
\label{sec:torus}
\begin{figure}[h!]
\epsscale{0.7}
\plotone{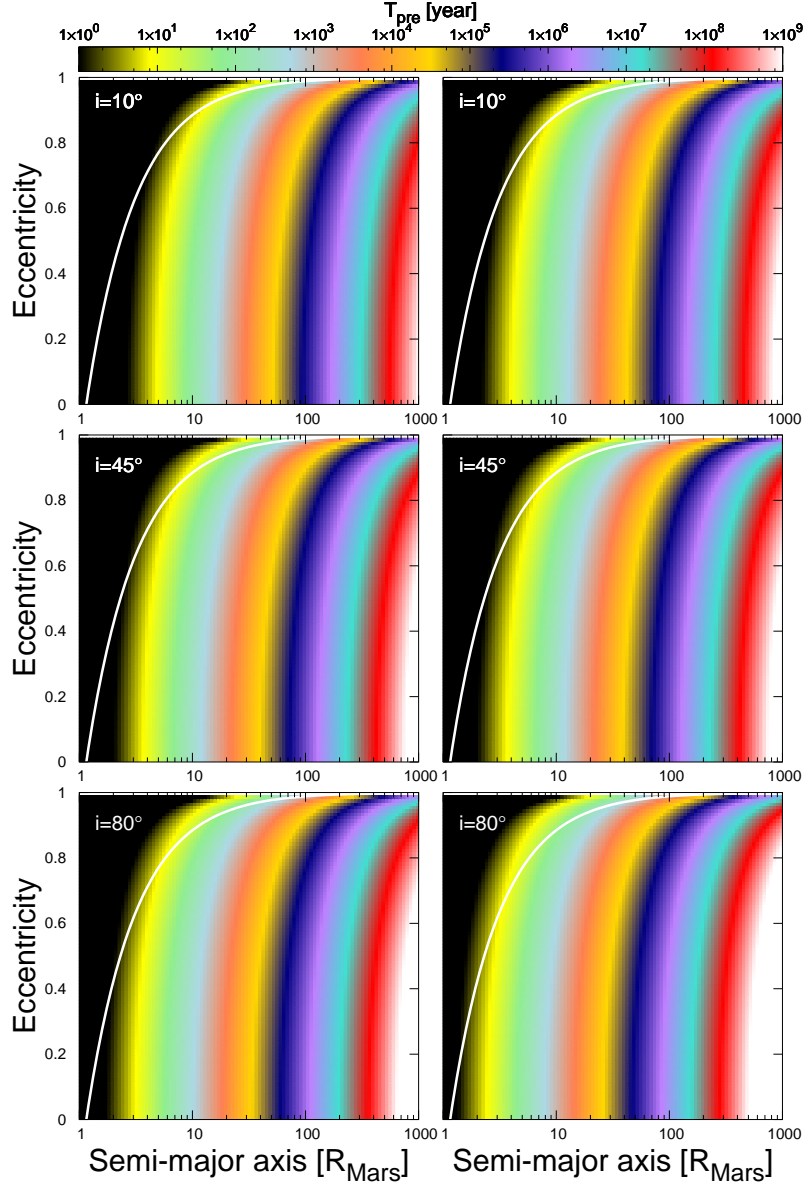}
\caption{Precession timescale $T_{\rm pre}$ as a function of semi-major axis and eccentricity at different disk inclination from the equatorial plane of Mars (from top to bottom panels, $i=10$, $45$ and $80$ degrees are shown, respectively). Left panel shows that for argument of pericenter ($T_{\rm \omega}$) and right panel shows that of longitude of the ascending node ($T_{\rm \Omega}$). White line shows the analytical $a-e$ distribution of the initial disk particles (equation \ref{e_a_analytic}). When $T_{\rm pre} < 1$ years, it is plotted with black and when $T_{\rm pre} > 10^{9}$ years, it is plotted with white.}
\label{map}
\end{figure}

Here, we will discuss the case when particles experience orbital precession and the system is randomized before particle-particle collisions are effective. As was also discussed in \cite{Hyo17c}, just after the impact, the particles have large eccentricities when they start to be influenced by Martian oblateness (mainly $J_2$ term), i.e. as their orbits (the argument of pericenter $\omega$ and the longitude of ascending node $\Omega$) start to precess around Mars. Using $J_{2}= 1.96 \times 10^{-3}$ value for Mars (\url{http://pds-geosciences.wustl.edu/mro/mro-m-rss-5-sdp-v1/mrors_1xxx/document/shadr.pdf}), we can calculate the timescales of precession as $T_{\omega}=2\pi / \dot{\omega}$ and $T_{\Omega}=2\pi/\dot{\Omega}$ where $\dot{\omega} = \frac{3n}{\left( 1-e^2 \right)^2} \left( \frac{R_{\rm Mars}}{a} \right)^2 \left( 1-\frac{5}{4}\sin^2(i) \right)J_2$ and $\dot{\Omega} =  -\frac{3n\cos(i)}{2\left( 1-e^2 \right)^2} \left( \frac{R_{\rm Mars}}{a} \right)^2 J_2$, respectively (Figure \ref{map}). Note that, the actual value of $J_2$ may differ from the current value and it is important to calculate the precession rate but this is the beyond the scope of our paper. The timescales depend on the eccentricity, semi-major axis and inclination from the equatorial plane of Mars but for parameters of interest here ($e \sim 0.5-0.9$ and $a \sim 2-10 R_{\rm Mars}$), timescales range from $1$ to $100$ years depending on the inclination of the disk with respect to the Martian equatorial plane (Figure \ref{map}). The ratio between these two precession timescales is $|T_{\omega}/T_{\Omega}| = |\frac{\cos(i)}{2(1-\frac{5}{4}\sin(i)^2)}|$ that only depends on $i$ and these timescales are comparable for the nominal case of $i=45$ degrees.\\

After the formation of a torus-like structure \citep[see also][]{Hyo17a,Hyo17c}, particle-particle collision may collape the system into equatorial plane. Here, we analytically estimate the collisional timescale between particles after they form a torus-like structure. In the particle-in-a-box approximation, the collision timescale can be written as 
\begin{equation}
	T_{\rm col} = \frac{1}{n_{\rm p}\sigma_{\rm col} v_{\rm rel}},
\label{tau_col}
\end{equation}
where $n_{\rm p}$ is the number density of particles, $\sigma_{\rm col}$ is the collision cross section and $v_{\rm rel}$ is the relative velocity. As also discussed in \cite{Hyo17a}, after the system forms a torus-like structure, orbits of particles are randomly oriented so that they cross each other and their relative velocity becomes about their local Keplerian velocity. Without gravitational focusing, the cross section can be written as $ \sigma_{\rm col} = 4\pi r_{\rm p}^2$, where $r_{\rm p}$ is the particle size. Here we can neglect the gravitational focusing term because, when particles are close to or within the Roche limit, gravitational attraction between particles becomes negligible \citep{Oht93,Hyo14,Hyo15b}. Following the argument of \cite{Hyo17a}, the volume number density of particles can be written as
\begin{equation}
	n_{\rm p}(r, \psi)= \int da \int de \int di \frac{ N(a,e,i) P(r| a,e) P(\psi| i)}{2 \pi r^2},
\label{number_density}
\end{equation}
where $N(a,e,i)$ is the number of particles as a function of $a$, $e$ and $i$. $P(r| a,e)$ and $P(\psi| i)$ are the probability of finding a particle at radial distance $r$ and an angle $\psi$ from the equatorial plane of the planet and are written as
\begin{equation}
	P(r| a,e)=\frac{r}{\pi a \sqrt{a^2e^2-(a-r)^2}},
\end{equation}
and 
\begin{equation}
	P(\psi| i)=\frac{|\cos(\psi)|}{\pi \sqrt{ \sin(i)^2 - \sin(\psi)^2}},
\end{equation}
respectively.\\

Here, in order to estimate the collision timescale, we first integrate $n_{\rm p}$ over $a$ and $e$ assuming the disk mean inclination of either $i=10$, $45$ and $80$ degrees. We use the data obtianed from SPH simulations from \cite{Hyo17c} (Figure \ref{ae_dis}) to know the distribution of $a$. Then, the distribution of $e$ is obtained using equation \ref{e_a_analytic}. \cite{Hyo17c} has considered that the particle size that formed during the impact ejection is regulated by their local shear velocity and material surface tension and they estimated that the typical particle size is  $r_{\rm p} \sim 1.5$ m. We assume particle density of $\rho_{\rm p}=2500$ kg m$^{-3}$ to calculate the number of particles as $N(a,e,i)=m_{\rm SPH}/m_{\rm p}$, where $m_{\rm SPH}$ is the mass of one SPH particle whose orbital elements are $a$ and $e$ and $m_{\rm p}=4\pi \rho_{\rm p} r_{\rm p}^3/3$ is the mass of particle. The collision timescale decreases when $r$ is close to the particles' pericenter and/or $\psi$ is closer to $i$. This is because near pericenter, apocenter and at maximum elevation, the radial and vertical orbital velocity is zero so that the particle residence time is at its maximum there. Thus, in this study, we assume that collision takes place at $r=1.1r_{\rm per}$ and $\psi = 0.9i$. Then, assuming the relative velocity of $v_{\rm rel} \sim v_{\rm peri} \sin(45^{\circ}$) for orbital elements of $e=0.8$, we estimate the collision timescale of less than $1\times 10^{-5}$ year for $i=10$, $45$ and $80$ degrees.  Note that, in the above argument, we assume arbitrary choices of the location of collision and relative velocity, but we confirm that the collision timescale is always much less than a year even with any other choises. The estimated collision timescales close to the pericenter are much smaller than the orbital period of particles. In contrast, the collision timescales at around apocenter distance are larger than the orbital period of particles.\\

Thus, as soon as the disk particles come back to their pericenter distances, they experience collisions. Therefore, after the formation of a torus-like structure, the system is expected to collapse through inelastic collisions to form a thin equatorial disk on a timescale comparable to their orbital period. In addition, the eccentric orbits of the debris are circularized at the same time when the inclination is damped, forming the inner dense and outer light radial profile of the disk in \cite{Ros16} under the assumption that every particle converges to the circular orbit corresponding to its angular momentum. However, we may need more detailed investigations that consider collisional fragmentation. We leave this matter for future works.\\

\subsubsection{The case for forming a flat inclined low-eccentricity disk}
\label{sec:flat_disk}
Even under the phase alignment, particles may experience high velocity collision at their pericenters because they share almost the same pericenter distances and have a wide distribution of large eccentricity (see Figures \ref{ae_dis} and \ref{phases}). In this subsection, we discuss the case when collisional damping at their pericenter is very efficient and thus a flat inclined low-eccentricity disk is formed. If collision occurs between particles who share the same pericenter distance $r_{\rm peri}$ but have different orbital elements ($e_{1,2}$ where $e_{1}>e_{2}$) under the phase aligment as seen in our case (see Figures \ref{ae_dis} and \ref{phases}), collision velocity can be estimated as 
\begin{equation}
	v_{\rm col, alig}=\sqrt{\frac{GM}{r_{\rm peri}}} \left(\sqrt{1+e_1} - \sqrt{1+e_2} \right) = 3.2 {\rm km \hspace{1mm}s}^{-1} \times \left( \frac{r}{3800 {\rm km}} \right)^{-1/2} \left( \frac{M}{6\times10^{23}{\rm kg}} \right)^{1/2} \left(\sqrt{1+e_1} - \sqrt{1+e_2} \right).
\end{equation}
Thus, if collision occurs between particles whose eccentricities are $e=0.9$ and $e=0$, respectively, we get $v_{\rm col, alig} \sim 1.2$ km s$^{-1}$. Then, their initial collision timescale is expected to be their synodic period when there is no dense circular inner rings as 
\begin{equation}
	T_{\rm syn} = 2\pi a /( \frac{3}{2}\Delta a \Omega) \sim 15 {\rm days} \times \left( \frac{M}{6\times10^{23} {\rm kg}} \right)^{-1/2} \left( \frac{\Delta a}{3400 {\rm km}} \right)^{-1} \left( \frac{a}{3.4 \times 10^4 {\rm km}} \right)^{5/2},
\label{Tsyn}
\end{equation}
where $\Delta a$ is the difference in semi-major axis of two colliding particles. In contrast, after dense circular rings are formed due to collision damping, the collision timescale between the dense circular rings and eccentric particles becomes about their Keplerian period ($\sim$ few days for particles whose $a=10R_{\rm Mars}$). Such high collision velocity may quickly damp the system and energetic enough to form  $\sim 100$ $\micron$ sized particles \citep{Hyo17c}. As also discussed in Section \ref{sec:torus}, the initial $a-e$ distribution is expected to be circularized through inelastic collisions while conserving the angular momentum of the disk, eventually forming the inner dense and outer light radial distribution of \cite{Ros16}. \\

The timescale of the circularization ($\sim$ tens of days, see equation \ref{Tsyn}) is much shorter than the precession timescale ($\sim 1-100$ years, see Section \ref{sec:torus}).  Therefore, the system forms an inclined low-eccentricity inner dense and outer light disk while the disk keeps its average inclination from the equatorial plane of Mars resulted directly from the giant impact, rather than the nodal precession forms a torus-like structure with large eccentricity.\\

After the formation of a flat inclined circular disk, neglecting the effect of self-gravity (we will discuss this effect in the next paragraph), the evolution is expected to be either of the following two extreme cases: (A) if collision timescale is  shorter than the differential precession timescale to form a torus-like structure, inelastic collision occurs between nearby particles whose longitudes of node are slightly different as a result of the differential nodal precession and their mean inclination decreases. This is because the nodal differential precession induced by $J_2$ term always tries to make the system symmetric to the equatorial plane of Mars. Thus, the disk takes gradual inside-out evolution by lowering its mean inclination to settle into the Martian equatorial plane with differential precession timescale (see also Figure \ref{map} at small $e$), or (B) if collision timescale is larger than the precession timescale, the system first forms a torus-like structure with the differential precession timescale. Then, collision will collapse the system into the equatorial plane. In our case, the collision timescale in the outer disk (in the case of a torus structure with its inclination $i=45$ degrees and $a=5.5R_{\rm Mars}$ with the radial width of $\Delta a=3R_{\rm Mars}$ where Phobos and Deimos are expected to form with a disk mass of $M_{\rm Phobos}$, and particle size of 1 m and its density 2500 kg m$^{-3}$) is estimated by using the same argument above (Equation \ref{tau_col}) where $n_{\rm p}=N/V$, $N=M_{\rm Phobos}/m_{\rm p} = 10^{16} {\rm kg}/10^4 {\rm kg} = 1 \times 10^{12}$ and $V=(2 \pi a) \Delta a (2a \tan(i)) \sim 4.4 \times 10^{22}$ m$^3$. Thus, $n_{\rm p} \sim 2.2 \times 10^{-11}$, $\sigma_{\rm col}=\pi (1.0{\rm m})^2 \sim 3$ m$^{2}$ and $v_{\rm rel} \sim V_{\rm Kep} \sin(i) = \sqrt{GM/a} \sin(i)=1000$ m s$^{-1}$. Thus, $T_{\rm col} \sim 0.5$ years. This timescale is actually the maximum timescale of collision because we assume the scale hight of $2 a \tan(i)$ and that the actual particle size is expected to be smaller due to collisional cascade. So, collision timescale is much smaller than the precession timescale. In the inner disk, the collision timescale is shorter than that of the outer disk due to its larger number density and larger relative velocity. Thus, in reality the case (A) occurs.\\

Lastly, using $N$-body simulations that include self-gravity between all particles, we investigate the evolution of the system by nodal precession. Under some condition, the self-gravity of the disk can prevent differential precession and the disk precesses rigidly without changing its inclination \citep{Bat12, Mor12}. Our $N$-body code is the same as that used in \cite{Hyo15a,Hyo17a} and \cite{Hyo17b}, and we include the effect of $J_2$ \citep[see more details][]{Hyo17a}. In the case of the canonical Martian-moon forming disk, the inner disk ($a<4 R_{\rm Mars}$) is massive ($\sim 10^{20}$ kg) and the outer disk ($4-7R_{\rm Mars}$) has a mass of Phobos ($\sim 10^{16}$ kg) \citep{Ros16,Hyo17c}. Orbits of the inner disk are expected to precess much quicker than those of the outer disk (we have confirmed by $N$-body simulations that the self-gravity within the inner disk does not prevent the differential precession and the case (A) discussed above is expected to occur on a timescale of $\sim$ few $10$ yeras (see Figure \ref{map} at $e=0$)). Thus, the inner part may quickly form a thin equatorial massive inner disk while the outer disk remains inclined. The orbital period of particles within the inner massive disk is much shorter than the precession timescale of the outer disk (Figure \ref{map}). Thus, we can approximate the effect of the massive inner disk on the nodal precession of the outer disk as a secular perturbation of the inner massive satellite \citep[see also][]{Mor12}. In addition, the inner massive disk quickly spreads and massive inner moons are formed \citep{CC12,Ros16}. Such inner massive disk or moon(s) increase the net effect of $J_2$ and the nodal precession can be accelerated. Thus, following the procedure in \cite{Mor12}, we perform $N$-body simulation including a single massive satellite (mass of $10^{20}$kg) on circular orbit ($a=2.5R_{\rm Mars}$) on the equatorial plane of Mars. In contrast, outer disk is represented by a swarm of 1000 equal-mass particles initially on circular orbits with their mean inclinations of $45$ degrees from the equatorial plane of Mars. In addition, we investigate the outer disk evolution without the inner massive satellite (which means without the effect of the inner disk).\\

\begin{figure}[ht!]
\epsscale{1.0}
\plotone{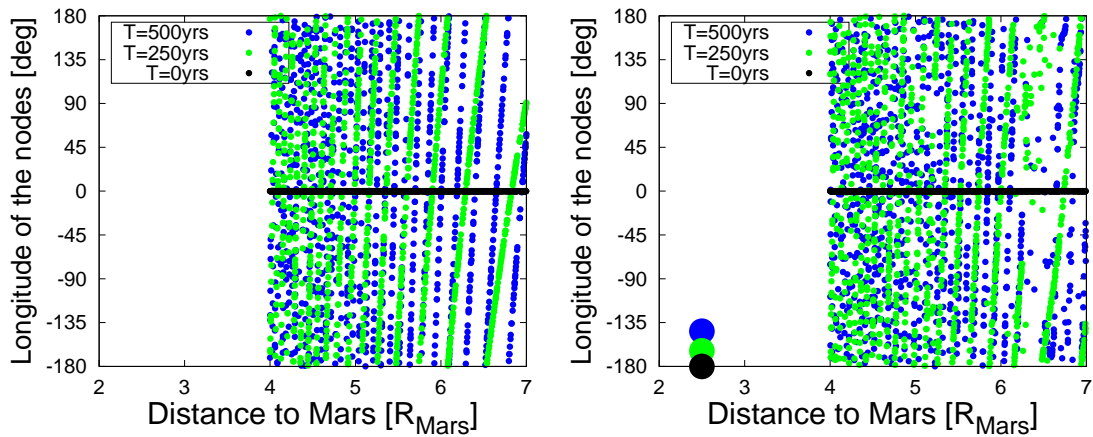}
\caption{The longitude of the nodes of disk particles under the effect of $J_2$ at different epoch obtained from our $N$-body simulations. Initially, particles on circular orbits are distributed between $4-7R_{\rm Mars}$ where Phobos and Deimos are expected to form \citep{Ros16}. Left panel shows the case of no inner massive satellite and right panel shows the case with a massive inner satellite. A large dot in the right panel shows a massive satellite. Black dots show outer disk particles at $T=0$ year, green dots shows those of $250$ years and blue dots show those after $500$ years.}
\label{Nbody}
\end{figure}

Figure \ref{Nbody} shows time evolutions of the longitude of nodes of disk particles. Inner satellite enhances the net effect of $J_2$ and the differential precession is slightly accelerated compared to the case of no inner massive disk \citep[see also][]{Mor12}. Our $N$-body simulations show that in the case of Phobos and Deimos forming disk ($\sim 10^{16}$ kg), the self-gravity is not effective to induce the rigid precession of the disk. In the framework of the giant impact hypothesis, the inner large moon is formed from the Roche-interior massive disk on a timescale of $\sim100$ years and migrate outward up to $4R_{\rm Mars}$ on a timescale of $\sim1000$ years \citep{Ros16}. As discussed above, in the outer disk, the collision timescale is shorter than the nodal precession timescale and thus the disk inclination decreases with the precession timescale (the case of (A)). Together with the results of $N$-body simulations (Figure \ref{Nbody}), the nodal precession timescale in the outer disk is few tens to few hundreds years (see Figure \ref{map}, $a=4-7R_{\rm Mars}$ and $e \sim 0$), depending on the initial inclination. Therefore, it is likely that the outer disk eventually settles into a thin equatorial near circular disk before the large inner moon migrates outward so that the equatorial Phobos and Deimos can accrete \citep{Ros16}.

 \section{Summary \& Discussion}\label{sec:summary}
The origin of Martian moons Phobos and Deimos is intensely debated. Recent works have shown that they might have accreted within a debris disk produced by the giant impact that formed the Borealis basin \citep{Ros16,Hes17}. If so, two dynamical questions naturally arise. First, why is the Borealis basin not on the Martian equatorial plane but close to the north pole? The Borealis basin forming impact can produce almost all of the current spin period of Mars. Thus, if there is no pre-impact Martian spin or a slow spin, the Borealis basin is expected to form around the equatorial plane and not on the northern hemisphere. Second, why do Phobos and Deimos orbit almost on the equatorial plane of Mars? If there is a pre-impact spin on Mars comparable to that given by the Borealis basin forming impact and if the pre-impact spin axis is not aligned to the angular momentum vector of the impactor, the resultant disk orbital plane is expected to be different from the equatorial plane of Mars and thus non-equatorial Phobos and Deimos may form.\\

In order to answer the first question -- why the Borealis basin is located close to the northern pole and not near the equatorial plane -- we have investigated the planetary reorientation due to the mass deficit at the Borealis basin (Section \ref{sec:tpw}). We have found that Borealis-induced True Polar Wander (TPW), using the equilibrium theory, can provide the required reorientation of the planet to move the center of the Borealis to its current latitudinal position from its initial location which is between 5$^\circ$N and 50$^\circ$N (more likely between 45$^\circ$N and 50$^\circ$N) for a lithospheric thickness between 50 km and 200 km (intermediate latitudes, between 5$^\circ$N and 45$^\circ$N are possible but for lithospheric thicknesses thinner than 50 km). However, this estimation could be refined by taking into account the post-Borealis TPW expected with large impact basins like Hellas or Utopia and with Tharsis, but this is out of the scope of this paper.\\

Based on the canonical Martian-moon-forming impact (an impact energy of $3 \times 10^{29}$ J and an impact angle of 45 degrees \citep{Ros16,Hyo17c}) which also formed the Borealis basin \citep{Mar08} and using analytical arguments, we have investigated the detailed post-impact disk. Just after the impact, disk particles have large eccentricities (Section \ref{subsec:a_e}) and almost the same pericenter distances. In addition, their orbits are almost aligned (“phase alignment”) at their longitude of the pericenters (Section \ref{sec:dynamics}) and the initial disk is expected to be inclined with respect to the equatorial plane of Mars. Thus, collisional damping is not efficient because collision velocity is only the order of their shear velocities except their pericenter distances where the maximum collision velocity is $\sim 1$ km s$^{-1}$. In this paper, we considered that the debris disk may experience either of the following two dynamical paths before forming a thin circular equatorial disk: formation of a torus-like structure (Section \ref{sec:torus}) or formation of a thin inclined (with respect to Mars' equatorial plane) circular disk  (Section \ref{sec:flat_disk}). Comparing the timescales of these two dynamical evolutions, we found that the formation of a thin inclined circular disk is expected to occur preferentially due to fast collisional damping at particle's orbit pericenter. Then, due to the differential nodal precession and particle-particle inelastic collisions, the inclined disk is expected to experience an inside-out evolution to gradually lower the mean inclination of the disk to eventually settle into the equatorial plane within 1000 years (Section \ref{sec:flat_disk}). Then, the thin equatorial circular disk is expected to form Phobos and Deimos near the equatorial plane \citep{Ros16}. Thus, the above arguments are likely to be a dynamical pathway that can answer our second question: why do Phobos and Deimos orbit almost on the equatorial plane of Mars?\\

As discussed above, the results and analytical arguments presented in this work have strengthened the giant impact origin of Phobos and Deimos. Together with the expected material properties of the building blocks of Phobos and Deimos (such as particle sizes, material provenance: Mars material or impactor material, and thermodynamic properties) that have also been investigated in the framework of the giant impact hypothesis \citep{Ron16,Hyo17c}, our results would finally be tested by a future sample return mission such as JAXA's Martian Moons eXploration (MMX) mission.\\

\appendix
\section{The equilibrium theory}
The equilibrium theory considers axisymmetric loads, thus allowing the computation of the polar wander in latitude alone. Although the Borealis basin is slightly elliptical, we will assume an axisymmetric load for direct application of the computation developed in the equilibrium theory. In this theory, planet tends to reorient in response to the modification of the planet inertia tensor induced by mass excess or deficit \citep{Mat06}. However, the rotational bulge mass excess counteracts this effect and the efficiency of the TPW depends on the ratio between mass excess/deficit and rotational bulge load that corresponds to the following $Q'$ coefficient:
\begin{equation}
	Q' = \frac{\frac{4\pi a^3 g}{5M}L'_{\rm 20}}{\frac{-1}{3\sqrt{5}}a^2 \Omega^2 k_{\rm f}^{T*}},
\label{Q_dash}
\end{equation}
where  $a$, $g$, $M$,  are, the equatorial radius ($3400$ km), the gravity ($3.711$ m s$^{-2}$), the mass ($6.4 \times10^{23}$ kg), and the rotation rate of Mars ($7.08  \times 10^{-5}$ rad s$^{-1}$), respectively, and $k_{\rm f}^{T*}$ is the tidal fluid Love number for a planet without a lithosphere (i.e. 1.1867, see Table 1 in  \cite{Mat06}). $L'_{20}$ is the second degree zonal term of the spherical expansion of the surface density of the axisymmetric mass deficit (so, it has a negative value,  \cite{Gol55}). It is given as:
\begin{equation}
	L'_{20} = 2 \pi \int_0^{\theta_0} \frac{3 \cos^2 \theta - 1}{2} \rho_{\rm s} (\theta) d\theta,
\label{L_dash}
\end{equation}
where $\theta$ is the co-latitude, $\theta_0$ is the co-latitudinal extent of the axisymmetric deficit load, and $\rho_{\rm s}$ its surface density. If we assume this surface density is constant and that the shape of the cavity can be approximated by a spherical cap, it becomes:
\begin{equation}
	L'_{20} =  -\pi  \cos( \theta_0)  \sin(\theta_0)^2  \rho h/2,
\label{L_dash2}
\end{equation}
where $\rho$ is the volume density of the excavated material (3000 kg m$^{-3}$) and $h$ is the height of the spherical cap, corresponding to the crustal dichotomy thickness (26 km, \citep{Neu04}).\\

The $\alpha$ coefficient in Eq. 1 is computed as in Eq. 26 of \cite{Mat06}:
\begin{equation}
	\alpha = \frac{1+k_f^L}{1-\frac{k_f^T}{k_f^{T*}}},
\label{L_dash2}
\end{equation}
where $k_f^T$ and $k_f^L$ are the fluid Tidal and load Love numbers for a planet with a lithosphere of a given thickness. The values of the Love numbers are extracted from Figure 3 of \cite{Mat10} and given in Table 1 as a function of lithospheric thickness. We considered a thin lithosphere as expected in early Mars history.\\

Considering the spherical triangle formed by the post-Tharsis TPW geographical pole, the pre-tharsis TPW geographical pole (or paleopole) and the center of Borealis, basic spherical trigonometric relationship yields: 
\begin{equation}
	\cos(\theta_{\rm L})=\cos(\theta_{\rm f})\cos(\delta) + \sin(\theta_{\rm f})\sin(\delta)\cos(\Delta L)
\end{equation}
with $\theta_{\rm L}$ and $\theta_{\rm f}$ are the initial and final co-latitudes of Borealis center, respectively. $\delta$ is the Tharsis-TPW displacement and $\Delta L$ is the longitude shift between Tharsis and Borealis central meridians. Given the Borealis final co-latitude $\theta_{\rm f} = 23^\circ$ (latitude of 67$^\circ$, \cite{And08}), the Tharsis TPW displacement  $\delta$ = 18.9$^\circ$ \citep{Mat10} and the longitude shift of 50$^\circ$ (see Section 2), it yields $\theta_{\rm L}  = 18^\circ$.

\begin{deluxetable*}{ccCrlc}[ht!]
\tablecaption{Love numbers of Mars and  parameter values for varying lithospheric thickness values}
\tablecolumns{4}
\tablenum{1}
\tablewidth{0pt}
\tablehead{
\colhead{Lithospheric thickness (km)} &
\colhead{Load Fluid Love number $k_f^L$} &
\colhead{Tidal fluid Love number $k_f^T$} &
\colhead{$\alpha$ parameter}
}
\startdata
50 & -0.9150 & 1.1100 & 1.2708\\
100 & -0.8529 &1.0370 & 1.1686\\
200 & -0.6906 & 0.8475& 1.0822\\
\enddata
\end{deluxetable*}

\acknowledgments
This work was supported by JSPS Grants-in-Aid for JSPS Fellows (17J01269) and by the JSPS-MAEDI bilateral joint research project (SAKURA program). PR was financially supported by the Belgian PRODEX program managed by the European Space Agency in collaboration with the Belgian Federal Science Policy Office. HG was financially supported by JSPS Grants-in-Aid for Scientific Research (15K13562, 17H02990, 17H06457). SC was supported by the UnivEarthS Labex programme at Sorbonne Paris Cite (ANR-10-LABX-0023 and ANR-11-IDEX-0005-02). Numerical computations were partly performed on the S-CAPAD platform, IPGP, France. Lastly, we thank Dr. Aur{\'e}lien Crida for comments that greatly improved the presentation of the manuscript.

\vspace{5mm}
\software{GADGET-2 (Springel 2005)}

\end{document}